\documentstyle[twoside,fleqn,espcrc2]{article}
\newcommand{\ttbs}{\char'134}
\newcommand{\AmS}{{\protect\the\textfont2
  A\kern-.1667em\lower.5ex\hbox{M}\kern-.125emS}}

\title{%
\begin{flushright}
{\small FUP-93-02, HUPD-9327, IPS-93-21, YAMAGATA-HEP-93-15}\\
\end{flushright}
Systematic study of autocorrelation time in pure SU(3)
       lattice gauge theory}

\author{ QCD\_TARO Collaboration \\
        K.Akemi
           \address{Computational Science Research Laboratory,
                    Fujitsu Limited, Mihama-ku, Chiba 261, Japan},
        Ph.de Forcrand
           \address{IPS,ETH-Z\"urich, CH-8092 Z\"urich, Switzerland},
        M.Fujisaki$^{\rm \ a}$,
        T.Hashimoto
           \address{Dept. of Applied Physics, Fukui University,
                    Fukui 910, Japan},
        S.Hioki
           \address{Dept. of Physics, Hiroshima University,
                    Higashi-Hiroshima 724, Japan},
        O.Miyamura$^{\rm \ d}$, \\
        A.Nakamura
           \address{Faculty of Education, Yamagata University,
                    Yamagata 990, Japan},
        M.Okuda$^{\rm \ a}$,
        I.O.Stamatescu
           \address{FESt Heidelberg and
                 Institut f\"ur Theoretische Physik, Universit\"at Heidelberg,
                    D-6900 Heidelberg, Germany},
        Y.Tago$^{\rm \ a}$
        and
        T.Takaishi$^{\rm \ d}$
       }

\begin{document}
\thispagestyle{empty}
\begin{abstract}

Results of our autocorrelation measurement performed on Fujitsu
AP1000 are reported.
We analyze (i) typical autocorrelation time,
(ii) optimal mixing ratio between overrelaxation and pseudo-heatbath
and (iii) critical behavior of autocorrelation time
around cross-over region with high
statistic in wide range of $\beta$ for pure SU(3) lattice
gauge theory on $8^4$, $16^4$ and $32^4$ lattices.
For the	mixing ratio K,  small value (3-7) looks optimal
in the confined region, and  reduces the integrated autocorrelation time
by a factor 2-4 compared to the pseudo-heatbath.
On the other hand in the deconfined phase,
correlation times are short, and overrelaxation does not seem to matter
For a fixed value of K(=9 in this paper),
the dynamical exponent of overrelaxation is consistent with 2

Autocorrelation measurement of the topological charge
on $32^3 \times 64$ lattice at $\beta$ = 6.0 is also briefly
mentioned.

\end{abstract}

\maketitle

\section{Aims of our study}

 Aims of our autocorrelation measurement are to investigate the followings.\\
1) typical autocorrelation time $\tau$\\
2) operator dependence of $\tau$\\
3) $\beta$ dependence of $\tau$\\
4) how overrelaxation(OR) reduces $\tau$ or not ?\\
5) optimal mixing ratio between OR and pseudo-heatbath\\
6) lattice size dependence of $\tau$\\
7) origin of large error of $\tau$\\
8) critical exponent of $\tau$\\
9) $\tau$ for topological object on large($32^3 \times 64$) lattice\\

Points 1 to 5 were previously reported upon at LAT92 \cite{Lat92}.
Points 6 - 8 are new, as well as some preliminary results about 9.

\section{Notations and Parameters}

Notations and parameters in this paper are as follows.\\
1) OBSERVABLE \\
  We adopt the $1\times 1$ wilson loop on $2^4$ lattice
  blocked from original lattice as an observable of our
  autocorrelation measurement.\\
2) MIXING PARAMETER $K$\\
  To check the efficiency of overrelaxation, we introduce the mixing
  parameter $K$ which means that  the ratio between
  the Brown-Woch microcanonical updating\cite{BW}
  and Cabibbo-Marinari pseudo heat bath updating\cite{CM} is set
  to be $K$ : 1.\\
3) AUTOCORRELATION FUNCTION\\
  The autocorrelation function $\rho(t)$ of observable
  $O$ ($=1\times 1$ wilson loop) is defined as:
\begin{equation}
\rho(t) = { {\langle(O(i)-O_A)(O(i+t)-O_B)\rangle} \over
      {\sqrt{\langle(O(i)-O_A)^2\rangle\langle(O(i+t)-O_B)^2\rangle}} }
\end{equation}
where $\langle\ \rangle$ denotes average over $i$ and
$O_A = \langle O(i)\rangle$ and $O_B = \langle O(i+t)\rangle$.\\
4) AUTOCORRELATION TIME\\
  The autocorrelation time is defined in this paper as:
\begin{equation}
 \tau_{int} = \rho(0) + 2 \sum_{t=1}^{N} \rho(t) {{N-t} \over N}
\end{equation}
where $N$ is determined so that $\tau_{int}$ is maximized,
but $N < 10$  \% of the total sample and $N < 3 \tau_{int}$.\\
5) LATTICE SIZE and $\beta$
\begin{description}
\item[$8^4$:] 29 $\beta$ values for $K$=9, 12 $\beta$ values for $K$=0 and
             4 $\beta$ values for $K$=0-20
\item[$16^4$:] 45 $\beta$s for $K$=9 and 2 $\beta$s for $K$=0-12
\item[$32^4$:] 4 $\beta$s for $K$=9
\end{description}

\begin{figure}[t]
\unitlength 1.0mm
\begin{picture}(60,40)(-10,-10)
\def\xw{60.000000} \def\yw{40.000000}
\put(-10,35){\large $\tau_{int}$}
\put(2,35){\large (a) $\beta$=5.70}
\put(2.727273,16.799999){\circle*{1.500000}}
\put(5.454545,15.200000){\circle*{1.500000}}
\put(10.909091,10.400000){\circle*{1.500000}}
\put(16.363636,10.400000){\circle*{1.500000}}
\put(21.818182,12.000000){\circle*{1.500000}}
\put(35.454544,26.400000){\circle*{1.500000}}
\put(43.636364,20.000000){\circle*{1.500000}}
\put(49.090908,21.600000){\circle*{1.500000}}
\put(57.272728,24.000000){\circle*{1.500000}}
\def\errorbar1#1#2#3{
\put(#1,#2){\line(0,1){#3}}
\put(#1,#2){\line(0,-1){#3}} }
\errorbar1{2.727273}{16.799999}{7.200000}
\errorbar1{5.454545}{15.200000}{4.000000}
\errorbar1{10.909091}{10.400000}{2.400000}
\errorbar1{16.363636}{10.400000}{0.800000}
\errorbar1{21.818182}{12.000000}{3.200000}
\errorbar1{35.454544}{26.400000}{4.800000}
\errorbar1{43.636364}{20.000000}{3.200000}
\errorbar1{49.090908}{21.600000}{4.000000}
\errorbar1{57.272728}{24.000000}{6.400000}
\put(2.727273,0){\line(0,1){1}}
\put(16.363636,0){\line(0,1){1}}
\put(30.000000,0){\line(0,1){1}}
\put(43.636364,0){\line(0,1){1}}
\put(57.272728,0){\line(0,1){1}}
\put(2.727273,0){\line(0,1){1}}
\put(0,0.000000){\line(1,0){1}}
\put(\xw,0.000000){\line(-1,0){1}}
\put(0,8.000000){\line(1,0){1}}
\put(\xw,8.000000){\line(-1,0){1}}
\put(0,16.000000){\line(1,0){1}}
\put(\xw,16.000000){\line(-1,0){1}}
\put(0,24.000000){\line(1,0){1}}
\put(\xw,24.000000){\line(-1,0){1}}
\put(0,32.000000){\line(1,0){1}}
\put(\xw,32.000000){\line(-1,0){1}}
\put(0,40.000000){\line(1,0){1}}
\put(\xw,40.000000){\line(-1,0){1}}
\put(0,0.000000){\line(1,0){1}}
\put(\xw,0.000000){\line(-1,0){1}}
\put(-3.900000,-1.500000){\large 0}
\put(-5.800000,6.500000){\large 10}
\put(-5.800000,14.500000){\large 20}
\put(-5.800000,22.500000){\large 30}
\put(-5.800000,30.500000){\large 40}
\put(-5.800000,38.500000){\large 50}
{\linethickness{0.25mm}
\put(  0,  0){\line(1,0){\xw}}
\put(  0,\yw){\line(1,0){\xw}}
\put(\xw,  0){\line(0,1){\yw}}
\put(  0,  0){\line(0,1){\yw}} }
\end{picture}
\unitlength 1.0mm
\begin{picture}(60,40)(-10,-10)
\def\xw{60.000000} \def\yw{40.000000}
\put(27,-8){\large $K$}
\put(-10,33){\large $\tau_{int}$}
\put(2,35){\large (b) $\beta$=6.15}
\put(-5,-14){\large Fig.1 $\tau_{int}$ vs $K$  on $8^4$}
\put(2.727273,16.000000){\circle*{1.500000}}
\put(5.454545,12.000000){\circle*{1.500000}}
\put(10.909091,18.000000){\circle*{1.500000}}
\put(16.363636,10.000000){\circle*{1.500000}}
\put(21.818182,20.000000){\circle*{1.500000}}
\put(35.454544,22.000000){\circle*{1.500000}}
\put(43.636364,20.000000){\circle*{1.500000}}
\put(49.090908,14.000000){\circle*{1.500000}}
\put(57.272728,20.000000){\circle*{1.500000}}
\def\errorbar1#1#2#3{
\put(#1,#2){\line(0,1){#3}}
\put(#1,#2){\line(0,-1){#3}} }
\errorbar1{2.727273}{16.000000}{6.000000}
\errorbar1{5.454545}{12.000000}{2.000000}
\errorbar1{10.909091}{18.000000}{8.000000}
\errorbar1{16.363636}{10.000000}{2.000000}
\errorbar1{21.818182}{20.000000}{6.000000}
\errorbar1{35.454544}{22.000000}{6.000000}
\errorbar1{43.636364}{20.000000}{10.000000}
\errorbar1{49.090908}{14.000000}{6.000000}
\errorbar1{57.272728}{20.000000}{6.000000}
\put(2.727273,0){\line(0,1){1}}
\put(16.363636,0){\line(0,1){1}}
\put(30.000000,0){\line(0,1){1}}
\put(43.636364,0){\line(0,1){1}}
\put(57.272728,0){\line(0,1){1}}
\put(2.727273,0){\line(0,1){1}}
\put(2.727273,-3.5){\large 0}
\put(16.363636,-3.5){\large 5}
\put(28.900000,-3.5){\large 10}
\put(42.536364,-3.5){\large 15}
\put(56.172728,-3.5){\large 20}
\put(0,0.000000){\line(1,0){1}}
\put(\xw,0.000000){\line(-1,0){1}}
\put(0,20.000000){\line(1,0){1}}
\put(\xw,20.000000){\line(-1,0){1}}
\put(0,40.000000){\line(1,0){1}}
\put(\xw,40.000000){\line(-1,0){1}}
\put(0,0.000000){\line(1,0){1}}
\put(\xw,0.000000){\line(-1,0){1}}
\put(-3.900000,-1.500000){\large 0}
\put(-3.900000, 8.500000){\large 5}
\put(-5.800000,18.500000){\large 10}
\put(-5.800000,28.500000){\large 15}
{\linethickness{0.25mm}
\put(  0,  0){\line(1,0){\xw}}
\put(  0,\yw){\line(1,0){\xw}}
\put(\xw,  0){\line(0,1){\yw}}
\put(  0,  0){\line(0,1){\yw}} }
\end{picture}
\end{figure}
\begin{figure}[t]
\unitlength 1.0mm
\begin{picture}(60,40)(-10,-10)
\def\xw{60.000000} \def\yw{40.000000}
\put(-10,35){\large $\tau_{int}$}
\put(2,35){\large (a) $\beta$=6.30}
\put(4.285714,17.299999){\circle*{1.500000}}
\put(17.142857,11.900000){\circle*{1.500000}}
\put(25.714285,6.800000){\circle*{1.500000}}
\put(34.285713,4.800000){\circle*{1.500000}}
\put(42.857143,27.600000){\circle*{1.500000}}
\put(55.714287,13.500000){\circle*{1.500000}}
\def\errorbar1#1#2#3{
\put(#1,#2){\line(0,1){#3}}
\put(#1,#2){\line(0,-1){#3}} }
\errorbar1{4.285714}{17.299999}{2.900000}
\errorbar1{17.142857}{11.900000}{5.100000}
\errorbar1{25.714285}{6.800000}{3.100000}
\errorbar1{34.285713}{4.800000}{0.300000}
\errorbar1{42.857143}{27.600000}{12.800000}
\errorbar1{55.714287}{13.500000}{5.500000}
\put(4.285714,0){\line(0,1){1}}
\put(17.142857,0){\line(0,1){1}}
\put(25.714285,0){\line(0,1){1}}
\put(34.285713,0){\line(0,1){1}}
\put(42.857143,0){\line(0,1){1}}
\put(55.714287,0){\line(0,1){1}}
\put(4.285714,0){\line(0,1){1}}
\put(0,0.000000){\line(1,0){1}}
\put(\xw,0.000000){\line(-1,0){1}}
\put(0,10.000000){\line(1,0){1}}
\put(\xw,10.000000){\line(-1,0){1}}
\put(0,20.000000){\line(1,0){1}}
\put(\xw,20.000000){\line(-1,0){1}}
\put(0,30.000000){\line(1,0){1}}
\put(\xw,30.000000){\line(-1,0){1}}
\put(0,40.000000){\line(1,0){1}}
\put(\xw,40.000000){\line(-1,0){1}}
\put(0,0.000000){\line(1,0){1}}
\put(\xw,0.000000){\line(-1,0){1}}
\put(-2.900000,-1.500000){\large 0}
\put(-6.700000,8.500000){\large 100}
\put(-6.700000,18.500000){\large 200}
\put(-6.700000,28.500000){\large 300}
\put(-6.700000,38.500000){\large 400}
{\linethickness{0.25mm}
\put(  0,  0){\line(1,0){\xw}}
\put(  0,\yw){\line(1,0){\xw}}
\put(\xw,  0){\line(0,1){\yw}}
\put(  0,  0){\line(0,1){\yw}} }
\end{picture}
\unitlength 1.0mm
\begin{picture}(60,40)(-10,-10)
\def\xw{60.000000} \def\yw{40.000000}
\put(27,-8){\large $K$}
\put(-10,36){\large $\tau_{int}$}
\put(2,35){\large (b) $\beta$=6.80}
\put(-5,-14){\large Fig.2 $\tau_{int}$ vs $K$  on $16^4$}
\put(4.285714,7.200000){\circle*{1.500000}}
\put(25.714285,26.799999){\circle*{1.500000}}
\put(34.285713,7.600000){\circle*{1.500000}}
\put(42.857143,8.800000){\circle*{1.500000}}
\put(55.714287,9.200000){\circle*{1.500000}}
\def\errorbar1#1#2#3{
\put(#1,#2){\line(0,1){#3}}
\put(#1,#2){\line(0,-1){#3}} }
\errorbar1{4.285714}{7.200000}{3.200000}
\errorbar1{25.714285}{26.799999}{8.800000}
\errorbar1{34.285713}{7.600000}{1.600000}
\errorbar1{42.857143}{8.800000}{1.600000}
\errorbar1{55.714287}{9.200000}{1.600000}
\put(4.285714,0){\line(0,1){1}}
\put(17.142857,0){\line(0,1){1}}
\put(25.714285,0){\line(0,1){1}}
\put(34.285713,0){\line(0,1){1}}
\put(42.857143,0){\line(0,1){1}}
\put(55.714287,0){\line(0,1){1}}
\put(4.285714,0){\line(0,1){1}}
\put(4.285714,-3.5){\large 0}
\put(17.142857,-3.5){\large 3}
\put(25.714285,-3.5){\large 5}
\put(34.285713,-3.5){\large 7}
\put(42.857143,-3.5){\large 9}
\put(54.614287,-3.5){\large 12}
\put(5.385714,-3.5){\large }
\put(0,0.000000){\line(1,0){1}}
\put(\xw,0.000000){\line(-1,0){1}}
\put(0,8.000000){\line(1,0){1}}
\put(\xw,8.000000){\line(-1,0){1}}
\put(0,16.000000){\line(1,0){1}}
\put(\xw,16.000000){\line(-1,0){1}}
\put(0,24.000000){\line(1,0){1}}
\put(\xw,24.000000){\line(-1,0){1}}
\put(0,32.000000){\line(1,0){1}}
\put(\xw,32.000000){\line(-1,0){1}}
\put(0,0.000000){\line(1,0){1}}
\put(\xw,0.000000){\line(-1,0){1}}
\put(-2.900000,-1.500000){\large 0}
\put(-4.800000,6.500000){\large 20}
\put(-4.800000,14.500000){\large 40}
\put(-4.800000,22.500000){\large 60}
\put(-4.800000,30.500000){\large 80}
{\linethickness{0.25mm}
\put(  0,  0){\line(1,0){\xw}}
\put(  0,\yw){\line(1,0){\xw}}
\put(\xw,  0){\line(0,1){\yw}}
\put(  0,  0){\line(0,1){\yw}} }
\end{picture}
\end{figure}

\section{RESULTS}

In Fig.1 and Fig.2 we show the mixing parameter dependence of the
autocorrelation time $\tau$ on $8^4$ and $16^4$ lattice respectively.
In both cases, 2 $\beta$ values are presented, just below the crossover
corresponding to confinement (a), and above it (b).
Note the expanded scale and short autocorrelation times in (b).
In the confined phase, a small value of K ( 3-7 ) seems to reduce the
autocorrelation time by a factor ~ 2-4 compared to pseudo-heatbath.
Above deconfinement, no clear K-dependence can be seen.

$\tau$ is affected by large errors in these figures.
To check the origin of this error, we prepare 40000 sweeps
on $16^4$ lattice at $\beta$=6.40 after the thermalization.
We divide 40000 sweeps into 4 bins of 10000 sweeps each.
Next we calculate $\rho(t)$ in each bin.
The result is shown in Fig.3.
We can see large discrepancy between the results obtained from these 4 bins.
Clearly 40000 sweeps are not sufficient to obtain precise values
for the autocorrelation function and time in this rather extreme case,
very close to the deconfinement transition.

Next we show the results of $\tau$ on different lattice sizes in Fig.3.
In this figure we plot $\tau$ on $8^4$, $16^4$ and $32^4$ lattices as a
function of the inverse lattice spacing $a^{-1}$.
In order to convert the lattice coupling $\beta (= 6/g^2)$ to the lattice
spacing $a$, we use results of ref.4, obtained by a Monte Carlo
Renormalization Group analysis.
{}From Fig.4 we conclude that the critical exponent $z$ defined by
\begin{equation}
( {\tau_L \over \tau_{\alpha L}} ) = ( {a_{\alpha L} \over a_L }) ^z
\end{equation}
is consistent with 2.
A detailed analysis is now in progress.

Finally we mention briefly our measurement of topological charge
on $32^3 \times 64$ lattice.
This analysis is performed on Fujitsu AP1000 with 1024 processors.
We prepare 830 configurations separated 100 updating sweeps at
$\beta$ = 6.00 on $32^3 \times 64$ lattice, and then perform blocking
twice to obtain blocked configurations of size $8^3 \times 16$.
Topological quantities are measured using the cooling method on $8^3 \times
16$.
Fig.5 shows the typical cooling history of topological charge $Q$, the sum of
the absolute value of the local topological density $I$, and the normalized
action $\tilde S$, rescaled so that an instanton configuration
has $\tilde S$ = 1. We use the value after 50 cooling sweeps as indicated
by the arrow
in Fig.5.
As for the integrated autocorrelation time of the topological charge,
$\tau$ turns out to be quite small compared to 100 sweeps which is the
interval between measurements.
In this case we have tried to fit $\rho(t)$ by the exponential function of
$t$ to get so called exponential autocorrelation time, but we can not
extract reliable value at this stage
since $\rho(t)$ is fluctuating around 0 from $t$ = 200.

\section*{ACKNOWLEDGEMENTS}

We are indebted to M.Ikesaka, Y.Inada, K.Inoue,
M.Ishii, T.Saito, T.Shimizu and H.Shiraishi
at the Fujitsu parallel computing research facilities
for their valuable comments on parallel computing.

\begin{figure}[t]
\unitlength 1.0mm
\begin{picture}(60,40)(-10,-10)
\def\xw{60.000000} \def\yw{40.000000}
\put(27,-8){\large $t$}
\put(-10,30){\large $\rho(t)$}
\put(-5,-14){\large Fig.3 $\rho(t)$ vs $t$ for 4 bins}
\put(0.000000,40.000000){\circle*{1.000000}}
\put(1.500000,36.977119){\circle*{1.000000}}
\put(3.000000,35.806019){\circle*{1.000000}}
\put(4.500000,34.306519){\circle*{1.000000}}
\put(6.000000,33.326241){\circle*{1.000000}}
\put(7.500000,32.731258){\circle*{1.000000}}
\put(9.000000,31.464439){\circle*{1.000000}}
\put(10.500000,31.505880){\circle*{1.000000}}
\put(12.000000,31.179239){\circle*{1.000000}}
\put(13.500000,30.189701){\circle*{1.000000}}
\put(15.000000,29.546440){\circle*{1.000000}}
\put(16.500000,29.076059){\circle*{1.000000}}
\put(18.000000,28.976419){\circle*{1.000000}}
\put(19.500000,27.793640){\circle*{1.000000}}
\put(21.000000,26.922400){\circle*{1.000000}}
\put(22.500000,25.690500){\circle*{1.000000}}
\put(24.000000,25.843800){\circle*{1.000000}}
\put(25.500000,24.443199){\circle*{1.000000}}
\put(27.000000,24.827801){\circle*{1.000000}}
\put(28.500000,24.756460){\circle*{1.000000}}
\put(30.000000,23.537161){\circle*{1.000000}}
\put(31.500000,24.682301){\circle*{1.000000}}
\put(33.000000,22.658180){\circle*{1.000000}}
\put(34.500000,22.582439){\circle*{1.000000}}
\put(36.000000,22.543840){\circle*{1.000000}}
\put(37.500000,24.350321){\circle*{1.000000}}
\put(39.000000,25.453720){\circle*{1.000000}}
\put(40.500000,25.449501){\circle*{1.000000}}
\put(42.000000,24.974380){\circle*{1.000000}}
\put(43.500000,24.099800){\circle*{1.000000}}
\put(45.000000,24.506439){\circle*{1.000000}}
\put(46.500000,22.985680){\circle*{1.000000}}
\put(48.000000,23.601540){\circle*{1.000000}}
\put(49.500000,23.188879){\circle*{1.000000}}
\put(51.000000,23.730820){\circle*{1.000000}}
\put(52.500000,24.283779){\circle*{1.000000}}
\put(54.000000,23.112480){\circle*{1.000000}}
\put(55.500000,22.641340){\circle*{1.000000}}
\put(57.000000,22.063419){\circle*{1.000000}}
\put(58.500000,22.480820){\circle*{1.000000}}
\put(60.000000,23.350121){\circle*{1.000000}}
\put(0.000000,40.000000){\circle{1.000000}}
\put(1.500000,34.920399){\circle{1.000000}}
\put(3.000000,32.421799){\circle{1.000000}}
\put(4.500000,31.075861){\circle{1.000000}}
\put(6.000000,29.512239){\circle{1.000000}}
\put(7.500000,28.584740){\circle{1.000000}}
\put(9.000000,27.380980){\circle{1.000000}}
\put(10.500000,26.140640){\circle{1.000000}}
\put(12.000000,24.225340){\circle{1.000000}}
\put(13.500000,22.128500){\circle{1.000000}}
\put(15.000000,23.828520){\circle{1.000000}}
\put(16.500000,23.215801){\circle{1.000000}}
\put(18.000000,23.573959){\circle{1.000000}}
\put(19.500000,24.624140){\circle{1.000000}}
\put(21.000000,22.634359){\circle{1.000000}}
\put(22.500000,20.380419){\circle{1.000000}}
\put(24.000000,21.381821){\circle{1.000000}}
\put(25.500000,18.080599){\circle{1.000000}}
\put(27.000000,15.730000){\circle{1.000000}}
\put(28.500000,12.487400){\circle{1.000000}}
\put(30.000000,8.927801){\circle{1.000000}}
\put(31.500000,11.195000){\circle{1.000000}}
\put(33.000000,12.714200){\circle{1.000000}}
\put(34.500000,17.958401){\circle{1.000000}}
\put(36.000000,9.816599){\circle{1.000000}}
\put(37.500000,3.227401){\circle{1.000000}}
\put(39.000000,8.681200){\circle{1.000000}}
\put(40.500000,11.223801){\circle{1.000000}}
\put(42.000000,18.027000){\circle{1.000000}}
\put(43.500000,18.354599){\circle{1.000000}}
\put(45.000000,18.968599){\circle{1.000000}}
\put(46.500000,16.815601){\circle{1.000000}}
\put(48.000000,16.203999){\circle{1.000000}}
\put(49.500000,11.516800){\circle{1.000000}}
\put(51.000000,17.822399){\circle{1.000000}}
\put(52.500000,16.334000){\circle{1.000000}}
\put(54.000000,17.089401){\circle{1.000000}}
\put(55.500000,12.786600){\circle{1.000000}}
\put(57.000000,8.554399){\circle{1.000000}}
\put(58.500000,16.981201){\circle{1.000000}}
\put(60.000000,13.157401){\circle{1.000000}}
\def\fullboxmark#1#2#3#4{
\multiput(#1,#2)(0,0.05){#3}{\line(1,0){#4}} }
\fullboxmark{-0.500000}{39.500000}{20}{1.000000}
\fullboxmark{1.000000}{35.086842}{20}{1.000000}
\fullboxmark{2.500000}{32.800541}{20}{1.000000}
\fullboxmark{4.000000}{31.512379}{20}{1.000000}
\fullboxmark{5.500000}{31.061340}{20}{1.000000}
\fullboxmark{7.000000}{30.449200}{20}{1.000000}
\fullboxmark{8.500000}{28.750681}{20}{1.000000}
\fullboxmark{10.000000}{27.484221}{20}{1.000000}
\fullboxmark{11.500000}{26.376720}{20}{1.000000}
\fullboxmark{13.000000}{25.802259}{20}{1.000000}
\fullboxmark{14.500000}{23.925980}{20}{1.000000}
\fullboxmark{16.000000}{22.914742}{20}{1.000000}
\fullboxmark{17.500000}{19.927019}{20}{1.000000}
\fullboxmark{19.000000}{21.645821}{20}{1.000000}
\fullboxmark{20.500000}{20.689220}{20}{1.000000}
\fullboxmark{22.000000}{20.120319}{20}{1.000000}
\fullboxmark{23.500000}{20.309839}{20}{1.000000}
\fullboxmark{25.000000}{20.664040}{20}{1.000000}
\fullboxmark{26.500000}{19.660320}{20}{1.000000}
\fullboxmark{28.000000}{19.162001}{20}{1.000000}
\fullboxmark{29.500000}{18.775999}{20}{1.000000}
\fullboxmark{31.000000}{20.304300}{20}{1.000000}
\fullboxmark{32.500000}{21.770981}{20}{1.000000}
\fullboxmark{34.000000}{22.321400}{20}{1.000000}
\fullboxmark{35.500000}{21.062981}{20}{1.000000}
\fullboxmark{37.000000}{20.521740}{20}{1.000000}
\fullboxmark{38.500000}{19.368401}{20}{1.000000}
\fullboxmark{40.000000}{20.002020}{20}{1.000000}
\fullboxmark{41.500000}{20.429981}{20}{1.000000}
\fullboxmark{43.000000}{21.585920}{20}{1.000000}
\fullboxmark{44.500000}{21.186661}{20}{1.000000}
\fullboxmark{46.000000}{20.557800}{20}{1.000000}
\fullboxmark{47.500000}{14.676800}{20}{1.000000}
\fullboxmark{49.000000}{18.038200}{20}{1.000000}
\fullboxmark{50.500000}{15.965799}{20}{1.000000}
\fullboxmark{52.000000}{13.226001}{20}{1.000000}
\fullboxmark{53.500000}{12.943201}{20}{1.000000}
\fullboxmark{55.000000}{12.711201}{20}{1.000000}
\fullboxmark{56.500000}{8.633601}{20}{1.000000}
\fullboxmark{58.000000}{6.298401}{20}{1.000000}
\fullboxmark{59.500000}{6.583199}{20}{1.000000}
\def\boxmark#1#2#3{
\multiput(#1,#2)(0,#3){2}{\line(1,0){#3}}
\multiput(#1,#2)(#3,0){2}{\line(0,1){#3}} }
\boxmark{-0.500000}{39.500000}{1.000000}
\boxmark{1.000000}{36.224720}{1.000000}
\boxmark{2.500000}{34.686520}{1.000000}
\boxmark{4.000000}{33.562038}{1.000000}
\boxmark{5.500000}{33.295101}{1.000000}
\boxmark{7.000000}{33.239700}{1.000000}
\boxmark{8.500000}{32.203140}{1.000000}
\boxmark{10.000000}{31.207880}{1.000000}
\boxmark{11.500000}{29.796619}{1.000000}
\boxmark{13.000000}{30.169920}{1.000000}
\boxmark{14.500000}{30.127960}{1.000000}
\boxmark{16.000000}{30.221899}{1.000000}
\boxmark{17.500000}{29.589220}{1.000000}
\boxmark{19.000000}{28.869820}{1.000000}
\boxmark{20.500000}{28.382700}{1.000000}
\boxmark{22.000000}{28.191061}{1.000000}
\boxmark{23.500000}{27.592701}{1.000000}
\boxmark{25.000000}{27.284840}{1.000000}
\boxmark{26.500000}{25.369678}{1.000000}
\boxmark{28.000000}{24.724499}{1.000000}
\boxmark{29.500000}{26.086899}{1.000000}
\boxmark{31.000000}{27.259100}{1.000000}
\boxmark{32.500000}{27.269939}{1.000000}
\boxmark{34.000000}{25.664879}{1.000000}
\boxmark{35.500000}{25.048920}{1.000000}
\boxmark{37.000000}{25.211281}{1.000000}
\boxmark{38.500000}{25.295160}{1.000000}
\boxmark{40.000000}{26.264299}{1.000000}
\boxmark{41.500000}{25.394220}{1.000000}
\boxmark{43.000000}{24.171240}{1.000000}
\boxmark{44.500000}{23.450239}{1.000000}
\boxmark{46.000000}{24.310341}{1.000000}
\boxmark{47.500000}{23.504801}{1.000000}
\boxmark{49.000000}{23.282841}{1.000000}
\boxmark{50.500000}{21.462420}{1.000000}
\boxmark{52.000000}{20.926661}{1.000000}
\boxmark{53.500000}{19.867519}{1.000000}
\boxmark{55.000000}{20.758101}{1.000000}
\boxmark{56.500000}{20.628180}{1.000000}
\boxmark{58.000000}{21.059780}{1.000000}
\boxmark{59.500000}{22.342781}{1.000000}
\put(0.000000,0){\line(0,1){1}}
\put(15.000000,0){\line(0,1){1}}
\put(30.000000,0){\line(0,1){1}}
\put(45.000000,0){\line(0,1){1}}
\put(60.000000,0){\line(0,1){1}}
\put(0.000000,0){\line(0,1){1}}
\put(0.000000,-3.5){\large 0}
\put(11.900000,-3.5){\large 100}
\put(26.900000,-3.5){\large 200}
\put(41.900000,-3.5){\large 300}
\put(56.900000,-3.5){\large 400}
\put(0,0.000000){\line(1,0){1}}
\put(\xw,0.000000){\line(-1,0){1}}
\put(0,20.000000){\line(1,0){1}}
\put(\xw,20.000000){\line(-1,0){1}}
\put(0,40.000000){\line(1,0){1}}
\put(\xw,40.000000){\line(-1,0){1}}
\put(0,40.000000){\line(1,0){1}}
\put(\xw,40.000000){\line(-1,0){1}}
\put(-8.800000,-1.500000){\large $10^{-2}$}
\put(-8.800000,18.500000){\large $10^{-1}$}
\put(-4.200000,38.500000){\large 0}
{\linethickness{0.25mm}
\put(  0,  0){\line(1,0){\xw}}
\put(  0,\yw){\line(1,0){\xw}}
\put(\xw,  0){\line(0,1){\yw}}
\put(  0,  0){\line(0,1){\yw}} }
\put(2.5,14){\circle*{1.000000}}
\put(2.5,10){\circle{1.000000}}
\fullboxmark{2}{6}{20}{1.000000}
\boxmark{2}{2}{1.000000}
\put(4,13){0-10000}
\put(4,9){10000-20000}
\put(4,5){20000-30000}
\put(4,1){30000-40000}
\end{picture}
\end{figure}

\begin{figure}[t]
\unitlength 1.0mm
\begin{picture}(60,40)(-10,-10)
\def\xw{60.000000} \def\yw{40.000000}
\put(16,-8){\large $a^{-1}$ (GeV)}
\put(-10,33){\large $\tau_{int}$}
\put(-5,-14){\large Fig.4 $\tau_{int}$ vs $a^{-1}$ for $8^4$,$16^4$ and $32^4$}
\def\errorbar1#1#2#3{
\put(#1,#2){\line(0,1){#3}}
\put(#1,#2){\line(0,-1){#3}} }
\errorbar1{7.858533}{6.361613}{2.006867}
\errorbar1{9.984600}{6.361613}{2.006867}
\errorbar1{11.832666}{8.027467}{3.180813}
\errorbar1{14.206400}{9.319600}{2.453173}
\errorbar1{16.225967}{12.723240}{2.006867}
\errorbar1{19.039200}{14.389067}{0.974187}
\errorbar1{21.380634}{15.281733}{2.163413}
\errorbar1{22.221033}{16.406000}{2.135573}
\errorbar1{23.017235}{17.629601}{1.116607}
\errorbar1{25.195002}{17.347067}{1.173941}
\errorbar1{25.864367}{18.639200}{1.173941}
\errorbar1{26.507700}{15.681200}{0.776704}
\errorbar1{28.323534}{18.156399}{1.017401}
\errorbar1{28.895267}{20.068666}{0.727631}
\errorbar1{29.452801}{21.643333}{1.116607}
\errorbar1{30.265968}{21.214134}{1.204907}
\errorbar1{31.052767}{22.652933}{1.053787}
\errorbar1{31.562868}{20.246799}{1.620253}
\errorbar1{35.149666}{13.885200}{1.064672}
\errorbar1{37.357670}{9.319600}{1.173941}
\errorbar1{38.000668}{9.319600}{1.173941}
\errorbar1{38.842667}{9.319600}{1.173941}
\errorbar1{39.257999}{11.267974}{1.701813}
\errorbar1{39.465668}{8.027467}{3.180813}
\errorbar1{39.878334}{11.267974}{0.832925}
\errorbar1{41.497002}{8.027467}{1.478987}
\errorbar1{43.467667}{10.375346}{2.006867}
\put(7.858533,6.361613){\circle*{1.500000}}
\put(9.984600,6.361613){\circle*{1.500000}}
\put(11.832666,8.027467){\circle*{1.500000}}
\put(14.206400,9.319600){\circle*{1.500000}}
\put(16.225967,12.723240){\circle*{1.500000}}
\put(19.039200,14.389067){\circle*{1.500000}}
\put(21.380634,15.281733){\circle*{1.500000}}
\put(22.221033,16.406000){\circle*{1.500000}}
\put(23.017235,17.629601){\circle*{1.500000}}
\put(25.195002,17.347067){\circle*{1.500000}}
\put(25.864367,18.639200){\circle*{1.500000}}
\put(26.507700,15.681200){\circle*{1.500000}}
\put(28.323534,18.156399){\circle*{1.500000}}
\put(28.895267,20.068666){\circle*{1.500000}}
\put(29.452801,21.643333){\circle*{1.500000}}
\put(30.265968,21.214134){\circle*{1.500000}}
\put(31.052767,22.652933){\circle*{1.500000}}
\put(31.562868,20.246799){\circle*{1.500000}}
\put(35.149666,13.885200){\circle*{1.500000}}
\put(37.357670,9.319600){\circle*{1.500000}}
\put(38.000668,9.319600){\circle*{1.500000}}
\put(38.842667,9.319600){\circle*{1.500000}}
\put(39.257999,11.267974){\circle*{1.500000}}
\put(39.465668,8.027467){\circle*{1.500000}}
\put(39.878334,11.267974){\circle*{1.500000}}
\put(41.497002,8.027467){\circle*{1.500000}}
\put(43.467667,10.375346){\circle*{1.500000}}
\def\errorbar1#1#2#3{
\put(#1,#2){\line(0,1){#3}}
\put(#1,#2){\line(0,-1){#3}} }
\errorbar1{27.437267}{8.027467}{5.633987}
\errorbar1{30.265968}{14.389067}{2.006867}
\errorbar1{31.052767}{17.898933}{1.339427}
\errorbar1{32.804832}{16.736933}{2.376987}
\errorbar1{33.997334}{24.082399}{1.098735}
\errorbar1{34.692001}{18.156399}{1.017401}
\errorbar1{35.149666}{20.587601}{2.069133}
\errorbar1{35.602001}{26.724266}{3.067253}
\errorbar1{36.268333}{20.419733}{0.684416}
\errorbar1{36.706333}{25.926399}{3.733974}
\errorbar1{37.357670}{29.241600}{2.569000}
\errorbar1{38.000668}{29.972935}{1.681480}
\errorbar1{38.422337}{29.166933}{3.231160}
\errorbar1{39.051003}{35.904667}{3.766800}
\errorbar1{39.465668}{34.335869}{2.783573}
\errorbar1{39.878334}{36.632133}{1.470707}
\errorbar1{40.083332}{36.032402}{3.119853}
\errorbar1{40.490669}{30.651335}{1.833960}
\errorbar1{41.094337}{37.296936}{5.443467}
\errorbar1{41.497002}{28.361334}{3.920106}
\errorbar1{41.596668}{27.112268}{6.302533}
\errorbar1{41.695335}{35.904667}{5.640707}
\errorbar1{41.895000}{30.962934}{4.561640}
\errorbar1{42.092335}{27.526134}{3.114627}
\errorbar1{42.487335}{23.804401}{1.253269}
\errorbar1{42.880669}{27.218533}{9.488307}
\errorbar1{43.467667}{24.172132}{3.742947}
\errorbar1{44.048336}{35.615601}{8.650920}
\errorbar1{44.819668}{21.063732}{3.387707}
\errorbar1{45.394333}{19.295467}{1.045240}
\errorbar1{45.960667}{22.042801}{3.095653}
\errorbar1{46.341331}{18.156399}{1.017401}
\errorbar1{46.715004}{18.866268}{2.090800}
\errorbar1{47.277004}{18.402802}{2.867493}
\errorbar1{48.208332}{16.406000}{1.388507}
\errorbar1{49.134003}{17.898933}{1.064672}
\errorbar1{50.051666}{20.419733}{1.209467}
\errorbar1{52.771000}{18.402802}{1.478987}
\put(27.437267,8.027467){\circle{1.500000}}
\put(30.265968,14.389067){\circle{1.500000}}
\put(31.052767,17.898933){\circle{1.500000}}
\put(32.804832,16.736933){\circle{1.500000}}
\put(33.997334,24.082399){\circle{1.500000}}
\put(34.692001,18.156399){\circle{1.500000}}
\put(35.149666,20.587601){\circle{1.500000}}
\put(35.602001,26.724266){\circle{1.500000}}
\put(36.268333,20.419733){\circle{1.500000}}
\put(36.706333,25.926399){\circle{1.500000}}
\put(37.357670,29.241600){\circle{1.500000}}
\put(38.000668,29.972935){\circle{1.500000}}
\put(38.422337,29.166933){\circle{1.500000}}
\put(39.051003,35.904667){\circle{1.500000}}
\put(39.465668,34.335869){\circle{1.500000}}
\put(39.878334,36.632133){\circle{1.500000}}
\put(40.083332,36.032402){\circle{1.500000}}
\put(40.490669,30.651335){\circle{1.500000}}
\put(41.094337,37.296936){\circle{1.500000}}
\put(41.497002,28.361334){\circle{1.500000}}
\put(41.596668,27.112268){\circle{1.500000}}
\put(41.695335,35.904667){\circle{1.500000}}
\put(41.895000,30.962934){\circle{1.500000}}
\put(42.092335,27.526134){\circle{1.500000}}
\put(42.487335,23.804401){\circle{1.500000}}
\put(42.880669,27.218533){\circle{1.500000}}
\put(43.467667,24.172132){\circle{1.500000}}
\put(44.048336,35.615601){\circle{1.500000}}
\put(44.819668,21.063732){\circle{1.500000}}
\put(45.394333,19.295467){\circle{1.500000}}
\put(45.960667,22.042801){\circle{1.500000}}
\put(46.341331,18.156399){\circle{1.500000}}
\put(46.715004,18.866268){\circle{1.500000}}
\put(47.277004,18.402802){\circle{1.500000}}
\put(48.208332,16.406000){\circle{1.500000}}
\put(49.134003,17.898933){\circle{1.500000}}
\put(50.051666,20.419733){\circle{1.500000}}
\put(52.771000,18.402802){\circle{1.500000}}
\def\errorbar1#1#2#3{
\put(#1,#2){\line(0,1){#3}}
\put(#1,#2){\line(0,-1){#3}} }
\errorbar1{40.490669}{27.476000}{2.275493}
\errorbar1{44.437000}{31.836666}{2.910440}
\errorbar1{46.341331}{33.718933}{5.575693}
\errorbar1{49.134003}{32.376801}{10.775373}
\errorbar1{52.771000}{38.864933}{2.287507}
\def\fullboxmark#1#2#3#4{
\multiput(#1,#2)(0,0.05){#3}{\line(1,0){#4}}
}\fullboxmark{39.740669}{26.726000}{30}{1.500000}
\fullboxmark{43.687000}{31.086666}{30}{1.500000}
\fullboxmark{45.591331}{32.968933}{30}{1.500000}
\fullboxmark{48.384003}{31.626801}{30}{1.500000}
\fullboxmark{52.021000}{38.114933}{30}{1.500000}
\put(12.000001,0){\line(0,1){1}}
\put(22.000002,0){\line(0,1){1}}
\put(32.000000,0){\line(0,1){1}}
\put(42.000000,0){\line(0,1){1}}
\put(52.000000,0){\line(0,1){1}}
\put(0.000000,0){\line(0,1){1}}
\put( 9,-3.5){\large 0.5}
\put(21,-3.5){\large 1}
\put(31,-3.5){\large 2}
\put(41,-3.5){\large 4}
\put(51,-3.5){\large 8}
\put(0,0.000000){\line(1,0){1}}
\put(\xw,0.000000){\line(-1,0){1}}
\put(0,13.333333){\line(1,0){1}}
\put(\xw,13.333333){\line(-1,0){1}}
\put(0,26.666666){\line(1,0){1}}
\put(\xw,26.666666){\line(-1,0){1}}
\put(0,40.000000){\line(1,0){1}}
\put(\xw,40.000000){\line(-1,0){1}}
\put(0,0.000000){\line(1,0){1}}
\put(\xw,0.000000){\line(-1,0){1}}
\put(-3.900000,-1.500000){\large 1}
\put(-5.900000,11.833333){\large $10^1$}
\put(-5.900000,25.166666){\large $10^2$}
\put(-5.900000,38.500000){\large $10^3$}
{\linethickness{0.25mm}
\put(  0,  0){\line(1,0){\xw}}
\put(  0,\yw){\line(1,0){\xw}}
\put(\xw,  0){\line(0,1){\yw}}
\put(  0,  0){\line(0,1){\yw}} }
\put(5,35){\circle*{1.500000}}
\put(5,30){\circle{1.500000}}
\fullboxmark{4.25}{24.25}{30}{1.500000}
\put(7,34){$8^4$}
\put(7,29){$16^4$}
\put(7,24){$32^4$}
\end{picture}
\end{figure}

\begin{figure}[t]
\unitlength 1.0mm
\begin{picture}(60,40)(-10,-10)
\def\xw{60.000000} \def\yw{40.000000}
\put(-5,-14){\large Fig.5 cooling history of $Q$,$I$ and $\tilde S$}
\put(27,-8){\large cooling sweeps}
\put(20.4,10){\vector(0,1){4}}
\put(0.400000,19.849581){\circle*{1.000000}}
\put(2.400000,17.648180){\circle*{1.000000}}
\put(4.400000,15.404720){\circle*{1.000000}}
\put(6.400000,15.319920){\circle*{1.000000}}
\put(8.400000,15.173580){\circle*{1.000000}}
\put(10.400000,15.491420){\circle*{1.000000}}
\put(12.400000,16.569380){\circle*{1.000000}}
\put(14.400000,16.563919){\circle*{1.000000}}
\put(16.400000,16.496559){\circle*{1.000000}}
\put(18.400000,16.436420){\circle*{1.000000}}
\put(20.400000,16.428579){\circle*{1.000000}}
\put(22.400000,16.432680){\circle*{1.000000}}
\put(24.400000,16.454979){\circle*{1.000000}}
\put(26.400000,16.472660){\circle*{1.000000}}
\put(28.400000,16.521540){\circle*{1.000000}}
\put(30.400000,16.538620){\circle*{1.000000}}
\put(32.400002,16.522800){\circle*{1.000000}}
\put(34.400002,16.532440){\circle*{1.000000}}
\put(36.400002,16.538681){\circle*{1.000000}}
\put(38.400002,16.544359){\circle*{1.000000}}
\put(40.400002,16.530180){\circle*{1.000000}}
\put(42.400002,16.530420){\circle*{1.000000}}
\put(44.400002,16.544701){\circle*{1.000000}}
\put(46.400002,16.537880){\circle*{1.000000}}
\put(48.400002,16.552179){\circle*{1.000000}}
\put(50.400002,16.572599){\circle*{1.000000}}
\put(52.400002,16.571760){\circle*{1.000000}}
\put(54.400002,16.610121){\circle*{1.000000}}
\put(56.400002,16.645460){\circle*{1.000000}}
\put(58.400002,16.685499){\circle*{1.000000}}
\put(8.400000,39.140778){\circle{1.000000}}
\put(10.400000,35.015339){\circle{1.000000}}
\put(12.400000,31.901560){\circle{1.000000}}
\put(14.400000,30.146580){\circle{1.000000}}
\put(16.400000,29.051680){\circle{1.000000}}
\put(18.400000,28.158539){\circle{1.000000}}
\put(20.400000,27.495279){\circle{1.000000}}
\put(22.400000,27.024660){\circle{1.000000}}
\put(24.400000,26.630459){\circle{1.000000}}
\put(26.400000,26.375540){\circle{1.000000}}
\put(28.400000,26.121679){\circle{1.000000}}
\put(30.400000,25.929001){\circle{1.000000}}
\put(32.400002,25.744841){\circle{1.000000}}
\put(34.400002,25.527340){\circle{1.000000}}
\put(36.400002,25.335520){\circle{1.000000}}
\put(38.400002,25.076120){\circle{1.000000}}
\put(40.400002,24.886181){\circle{1.000000}}
\put(42.400002,24.680580){\circle{1.000000}}
\put(44.400002,24.509819){\circle{1.000000}}
\put(46.400002,24.382801){\circle{1.000000}}
\put(48.400002,24.237020){\circle{1.000000}}
\put(50.400002,24.144979){\circle{1.000000}}
\put(52.400002,24.069139){\circle{1.000000}}
\put(54.400002,23.988041){\circle{1.000000}}
\put(56.400002,23.917500){\circle{1.000000}}
\put(58.400002,23.863279){\circle{1.000000}}
\def\fullboxmark#1#2#3#4{
\multiput(#1,#2)(0,0.05){#3}{\line(1,0){#4}}
}\fullboxmark{9.900000}{36.861038}{20}{1.000000}
\fullboxmark{11.900000}{33.331100}{20}{1.000000}
\fullboxmark{13.900000}{30.734661}{20}{1.000000}
\fullboxmark{15.900000}{28.957439}{20}{1.000000}
\fullboxmark{17.900000}{27.624840}{20}{1.000000}
\fullboxmark{19.900000}{26.682020}{20}{1.000000}
\fullboxmark{21.900000}{26.066780}{20}{1.000000}
\fullboxmark{23.900000}{25.663919}{20}{1.000000}
\fullboxmark{25.900000}{25.398840}{20}{1.000000}
\fullboxmark{27.900000}{25.195080}{20}{1.000000}
\fullboxmark{29.900000}{25.046801}{20}{1.000000}
\fullboxmark{31.900002}{24.911360}{20}{1.000000}
\fullboxmark{33.900002}{24.776360}{20}{1.000000}
\fullboxmark{35.900002}{24.640680}{20}{1.000000}
\fullboxmark{37.900002}{24.503120}{20}{1.000000}
\fullboxmark{39.900002}{24.369780}{20}{1.000000}
\fullboxmark{41.900002}{24.230360}{20}{1.000000}
\fullboxmark{43.900002}{24.108419}{20}{1.000000}
\fullboxmark{45.900002}{23.994659}{20}{1.000000}
\fullboxmark{47.900002}{23.882040}{20}{1.000000}
\fullboxmark{49.900002}{23.792301}{20}{1.000000}
\fullboxmark{51.900002}{23.695080}{20}{1.000000}
\fullboxmark{53.900002}{23.618599}{20}{1.000000}
\fullboxmark{55.900002}{23.543301}{20}{1.000000}
\fullboxmark{57.900002}{23.490841}{20}{1.000000}
\put(0.000000,0){\line(0,1){1}}
\put(20.000000,0){\line(0,1){1}}
\put(40.000000,0){\line(0,1){1}}
\put(60.000000,0){\line(0,1){1}}
\put(0.000000,0){\line(0,1){1}}
\put(0.000000,-3.5){\large 0}
\put(18.900000,-3.5){\large 50}
\put(37.800000,-3.5){\large 100}
\put(57.800000,-3.5){\large 150}
\put(1.100000,-3.5){\large }
\put(0,4.000000){\line(1,0){1}}
\put(\xw,4.000000){\line(-1,0){1}}
\put(0,8.000000){\line(1,0){1}}
\put(\xw,8.000000){\line(-1,0){1}}
\put(0,12.000000){\line(1,0){1}}
\put(\xw,12.000000){\line(-1,0){1}}
\put(0,16.000000){\line(1,0){1}}
\put(\xw,16.000000){\line(-1,0){1}}
\put(0,20.000000){\line(1,0){1}}
\put(\xw,20.000000){\line(-1,0){1}}
\put(0,24.000000){\line(1,0){1}}
\put(\xw,24.000000){\line(-1,0){1}}
\put(0,28.000000){\line(1,0){1}}
\put(\xw,28.000000){\line(-1,0){1}}
\put(0,32.000000){\line(1,0){1}}
\put(\xw,32.000000){\line(-1,0){1}}
\put(0,36.000000){\line(1,0){1}}
\put(\xw,36.000000){\line(-1,0){1}}
\put(0,20.000000){\line(1,0){1}}
\put(\xw,20.000000){\line(-1,0){1}}
\put(-5.800000,2.500000){\large -8}
\put(-5.800000,10.500000){\large -4}
\put(-3.900000,18.500000){\large 0}
\put(-3.900000,26.500000){\large 4}
\put(-3.900000,34.500000){\large 8}
{\linethickness{0.25mm}
\put(  0,  0){\line(1,0){\xw}}
\put(  0,\yw){\line(1,0){\xw}}
\put(\xw,  0){\line(0,1){\yw}}
\put(  0,  0){\line(0,1){\yw}} }
\put(40.5,11){\circle*{1.000000}}
\put(40.5,7){\circle{1.000000}}
\fullboxmark{40}{3}{20}{1.000000}
\put(42,10){$Q$}
\put(42,6){$I$}
\put(42,2){$\tilde S$}
\end{picture}
\end{figure}

\end{document}